\def\@aabuffer{}
\def\author #1{\expandafter\def\expandafter\@aabuffer\expandafter
{\@aabuffer \small\rm      #1\relax \par}}
\def\address#1{\expandafter\def\expandafter\@aabuffer\expandafter
{\@aabuffer \small\it #1\relax \par\vspace{1em}}}
\def\maketitle{
\begin{center}
   {\bf \@title \par}       
   \vskip 2em                   
   \@aabuffer\relax
\end{center} \par
\gdef\@aabuffer{}
}
\def\abstracts#1{
\begin{center}
{\begin{minipage}{4.2truein}
                 \footnotesize
                 \parindent=0pt #1\par
                 \end{minipage}}\end{center}
                 \vskip 2em \par}
\fussy
\def\baselinestretch{1.0}

\ifx\selectfont\undefined
\def\@singlespacing{%
\def\baselinestretch{1}\ifx\@currsize\normalsize\@normalsize\else\@currsize\fi%
}
\else
\def\@singlespacing{\def\baselinestretch{1}\let\glb@currsize=\relax\selectfont}
\fi

\long\def\@makecaption#1#2{
   \vskip 10pt 
   \setbox\@tempboxa\hbox{\footnotesize\rm #1: #2}
   \ifdim \wd\@tempboxa >\hsize   
       \leftskip 0pt plus 1fil 
       \rightskip 0pt plus -1fil 
       \parfillskip 0pt plus 2fil 
       \footnotesize\rm #1: #2\par   
     \else                        
       \hbox to\hsize{\hfil\box\@tempboxa\hfil}  
   \fi}

\documentstyle[sprocl,epsf]{article}

\def\Journal#1#2#3#4{{#1} {\bf #2}, #3 (#4)}
\def\NPB{{\em Nucl. Phys.} B}
\def\PLB{{\em Phys. Lett.}  B}
\def\PRL{\em Phys. Rev. Lett.}
\def\PRD{{\em Phys. Rev.} D}
\def\VEV#1{\left\langle #1\right\rangle}
\def\ph#1{hep-ph/#1}
\def\etal{{\it et. al.} }

\begin{document}

\title{POLARIZED PARTON DISTRIBUTIONS FOR SPIN ASYMMETRIES}

\author{G. P. RAMSEY}

\address{Loyola University of Chicago, 6525 N. Sheridan, 
Chicago,\\ IL 60626, USA\\E-mail: gpr@hep.anl.gov} 

\maketitle\abstracts{We have used QCD and polarized deep-inelastic scattering
data to construct x-dependent polarized parton distributions. These flavor
dependent distributions evolve under the NLO DGLAP equations, satisfy positivity
constraints and agree well with the data. We perform all of the analysis in
x-space, avoiding difficulties with moments, and use these distributions to
predict spin observables. The small-x behavior is sensitive to the model for
the polarized gluon distribution.}

Polarized $x$-dependent parton distributions can be used to predict spin 
observables, such as hard scattering cross sections and asymmetries for
polarized processes.\cite{ggr} We have created flavor dependent distributions
at $Q_0^2=1$ GeV$^2$, motivated by physical constraints and data, then evolved
them to arbitrary $Q^2$ to predict polarized observables. The polarized 
distributions are generated from the unpolarized ones and theoretical
assumptions.

The distributions are constructed subject to the following: \\
(1) the integration over $x$ should reproduce the values extracted from
PDIS data,\cite{data} so that the fraction of spin carried by each constituent
is contained implicitly in the flavor-dependent distributions \\
(2) they should reproduce the $x$-dependent polarized structure
functions, $g_1^i$, $i$=p,n and d at the average $Q^2$ values of the data \\
(3) the small-$x$ behavior of $g_1(x)$ should fall between a Regge-like
 power of $x$ and a gluon-dominated logarithmic behavior \\
(4) the $Q^2$ behavior is generated by the NLO evolution equations, \\
(5) the positivity constraints are satisfied for all of the flavors. \\

The first two constraints build in compatibility with the PDIS data. The
third and fourth satisfy the appropriate $x$ and $Q^2$ kinematical dependence.
The last constraint is fundamental to our physical interpretation of
polarization.

The polarized valence distributions are determined from the unpolarized
distributions by imposing a modified SU(6) model,\cite{gr,qrrs} which
satisfies the Bjorken Sum Rule (BSR) with QCD corrections.

For the sea, we use a broken SU(3) model, to account for mass effects in 
polarizing strange quarks. We include charm via the evolution equations
($N_f$), at the appropriate $Q^2$ of charm production. The entire NLO 
analysis is done in $x$-space. A basic sea model, \cite{qrrs} and data
motivate the following form used for each sea flavor:
$\Delta q_f(x)\equiv \eta_f(x)\>x\>q_f(x)$, where $q_f(x)$ is the unpolarized
distribution for flavor $f$.\cite{cteq} Then, $\eta_f(x)$ is chosen to satisfy
the normalization:
\begin{equation}
\VEV{\Delta q_f}=\int_0^1 \eta_f(x)\>x\>q_f(x)\>dx\equiv \VEV{\eta xq_f}
=\eta_{av} \VEV{xq_f},  \label{eq:s1}
\end{equation}
where $\eta_{av}$ is extracted from data for each flavor. \cite{gr} Physically,
$\eta(x)$ may be interpreted as a modification of $\Delta q$ due to effects of
soft physics at low $x$.

The gauge- (or $\overline{MS}$) and chiral-invariant (or AB) schemes
representing the polarized sea are related by: \cite{cheng}
$\Delta q(x)_{GI}=\Delta q(x)_{CI}-(N_f\alpha_s/{2\pi})\Delta G(x)$,
where the GI and CI refer to the gauge- and chiral-invariant schemes,
respectively. Since there exists no empirical evidence that $\Delta G$
is very large at the relatively small $Q^2$ values of the data, \cite{e704}
we assume three different moderate $\Delta G$ models. The first model assumes
a positively polarized glue: $\Delta G(x)=xG(x).$ In the second model, we set
$\Delta G=0$, which is equivalent to the gauge-invariant scheme, since the
anomaly term vanishes. The third model is motivated by an instanton-induced
$\Delta G$, which is negatively polarized at small-$x$. \cite{koch}

\begin{figure}[t]
\epsfxsize=7cm
\centerline{\epsfbox{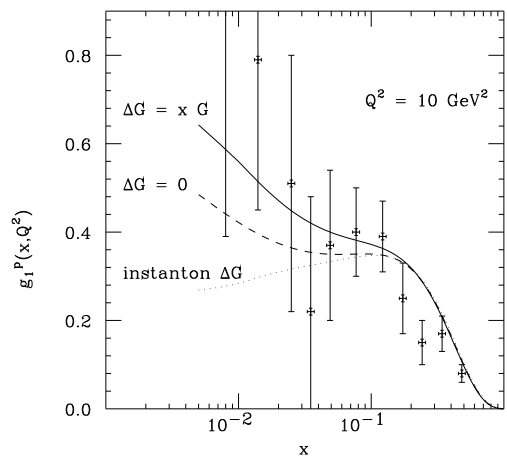}}
\caption{$g^p_1(x)$ for three models of $\Delta G$ compared to SMC data.
 \label{fig:g1p}}
\end{figure}

The overall parametrization for each of the polarized sea flavors, including
all gluon models is given in reference 1. We have used these to calculate the
structure functions, $g_1^{p,n,D}(x)$. These all agree with data at the
average $Q^2$ value for each set.\cite{ggr,data} The plot for $g_1^p$ is
shown in Fig. \ref{fig:g1p}.

The polarized sea distributions exhibit Regge-like behavior at small-$x$ and
become slightly positive at moderate $x$. With the positive gluon model, we
find that $g_1^p\sim x^{-0.19}$ at small $x$. It does not have the steep rise 
characteristic of the logarithmic behavior due to gluon exchange,\cite{cr} but
is compatible with Regge behavior. The distributions are consistent with data
at small-$x$ and are compatible with the appropriate counting rules at
large-$x$. \cite{bbs}

It may be possible to narrow down the models with more precise PDIS
experiments at small-$x$, which are planned at SLAC (E155) and DESY (HERMES).
However, there are other experiments which would provide a better indication
of the size and shape of $\Delta G$, including: (1) jet production in $e-p$
and $p-p$ collisions, (2) prompt photon production, (3) charm production and
(4) meson production. Groups at RHIC, CERN and DESY are planning these
experiments. For details, see reference 11. We are presently calculating jet
and direct photon production asymmetries using these distributions.\cite{goram}

\section*{Acknowledgments}
This work was done with L. E. Gordon (Jefferson Labs) and M. Goshtasbpour
(Shahid Beheshti Univ., Iran). See their articles in these proceedings.

\end{document}